\documentclass[pra,showpacs]{revtex4}
 \usepackage{amssymb} \usepackage{graphicx} \graphicspath{ {images/} }
  \usepackage{dcolumn} \usepackage{amsmath}
 \usepackage{tikz}
  \usetikzlibrary{shapes,backgrounds}
\usepackage{color}
\usepackage{varioref}

\begin{document}
\title{Spin-1/2 and spin-3/2 field solutions in plane wave spacetimes}
 \author{\"{O}zg\"{u}r A\c{c}{\i}k}
\email{ozacik@science.ankara.edu.tr}
\address{Department of Physics,
Ankara University, Faculty of Sciences, 06100, Tando\u gan-Ankara,
Turkey\\}

\begin{abstract}
We have found two non-trivial massless Dirac and two massive Rarita-Schwinger solutions in plane wave spacetimes. The first order symmetry operator transforming one of the massless Dirac solution to the other is constructed. The only non-vanishing spinor bilinear generated by the standard spinor basis is obtained and algebraic relations between the induced parallel forms are demonstrated. It is also seen that the spin-3/2 norm of the Rarita-Schwinger solutions enforces to the massless sector.

\end{abstract}
\def\nblx#1{\nabla_{X_{#1}}}
\def\sbl#1{(\psi\overline{\psi})_{#1}}

\maketitle
\section{Introduction}
Propagating solutions of the gravitational field equations of general relativity are termed as gravitational wave spacetimes. This class of spacetimes include plane fronted waves which have many interesting collision properties \cite{Griffiths}. They carry non-zero energy and momentum and describe the helicity-2 graviton states \cite{Dereli and Tucker}; the scattering of spinning classical test particles from them is also a vast subject of its own and can be thought as a continuation of the era opened by Mathisson, Fock, Papapetrou, Dixon and others \cite{Mohseni and Tucker}. The propagation of Dirac fields have been analyzed in this class and solutions to the Dirac equation have been found \cite{Talebaoui}. This suggests the uncovering of other solutions by considering the first order symmetry operators in curved spacetimes \cite{Acik Ertem Onder and Vercin} which depend on the existence of the hidden symmetries of spacetime \cite{Acik Ertem Onder Vercin1}. The solutions of Dirac equations are considered as if the Dirac fields are taken as test fields and their reaction on the curvature are neglected; this is because the Einstein-Dirac coupled field equations are problematic in many cases \cite{Radford and Clotz}. The plane wave Dirac solutions appear in two categories, massless \cite{Benn and Tucker} and massive \cite{Talebaoui}. Both categories are based on the existence of a specific kind of spinors, namely parallel spinors that are defined with respect to the spinor connection over the spinor bundle induced from the ordinary Levi-Civita connection on spacetime. These mentioned constructions are easily generated by functional linear combinations or functional multiples of the parallel spinors, where the latter function includes a mass factor for generating the massive sector.

Einstein-Rarita-Schwinger coupling is also problematic in many respects both in classical and quantum levels. For example the minimal coupling of Rarita-Schwinger field to a background gravitational field leads to a constraint which gives a trivial solution if the spacetime is not an Einstein space, i.e. the Ricci tensor is not proportional to the metric tensor field. The quantum theory of the spin-3/2 field minimally coupled to the electromagnetic field violates unitarity \cite{Hack and Makedonski}. Although a consistent quantization of spin-3/2 fields in gravitational fields seems to be background dependent \cite{Hack and Makedonski}, a class of spacetimes like Friedmann-Robertson-Walker type admits the consistency by the help of a positive definite inner product generating current \cite{Schenkel and Uhlemann}. Another interesting subject is the spin raising and lowering procedure in curved spacetimes based on the presence of twistor spinors \cite{Penrose Rindler II, Charlton}, where it can be shown that a free Maxwell field can be spin raised to a massless Rarita-Schwinger field by a twistor \cite{Acik Ertem 2017}. This procedure is a direct generalization for constructing massless Dirac solutions by using twistors \cite{Benn Kress spin}. Twistor spinors are also in the forefront of theoretical physics because of their usage for calculating scattering amplitudes in string theory \cite{Cachazo and Svrcek}. It is also known that conserved charges defined for spin-3/2 fields constitute a twistor space \cite{Penrose 3/2}.

In the present work we focus our attention to the analysis of massless Dirac spinors and massive spin-3/2 fields and solve the corresponding field equations in a class of plane fronted gravitational wave spacetimes. Our methods will be based on the coordinate free techniques gathered in the Clifford bundle formalism of differential forms and spinors \cite{Benn and Tucker, Chevalley, Lawson and Michelsohn, Hijazi et al} rather than the component oriented formalism of spinorial techniques \cite{Penrose and Newmann, Penrose Rindler I, Penrose Rindler II}. For these reasons we apply coordinate free pseudo-Riemannian differential geometry and spin geometry techniques throughout the paper and give a brief review for the tensor spinors in the appendix. The Rarita-Schwinger solutions in this work will be directly solved without any dependence on the existence of twistors.

The paper is organized as follows. In Section 2 we construct specific bases and co-bases for the plane wave spacetime class that we are considering. We also obtain the curvature characteristics by first calculating the Levi-Civita connection one forms. In Section 3 we pass to the evaluation of the spinor basis and the related symplectic inner product. In Section 4, from the manipulation of the spinorial covariant derivatives we show that there are two explicit parallel spinors inside the basis. Section 5 is devoted to the main result of the paper, namely to the two massless Dirac and two Rarita-Schwinger solutions. In Section 6 we deal with symmetry, mapping one of the massless Dirac solution to the other one. Section 7 concludes the paper with a following appendix containing the geometry and algebra of tensor spinors.
\section{Some basis for plane wave spacetimes}
We consider plane wave spacetime $(M,g)$ with $g$ given locally in $(u,v,x,y)$ coordinate chart by
\begin{eqnarray}
g=2\big(du\otimes dv+ dv \otimes du-2Hdu\otimes du+dz\otimes dz^{*}+dz^{*} \otimes dz\big)
\end{eqnarray}
where $z=x+\mathrm{i}y$ and $H=H(u,z,z^{*})$. It is possible to adopt a null co-basis $\{n^{a}\}  ;a=1,2,3,4$
\begin{eqnarray*}
  n^{1}&=&du \\
  n^{2}&=&dv-Hdu \\
  n^{3}&=&dz \\
  n^{4}&=&dz^{*}
\end{eqnarray*}
and its dual $\{Y_{b}\}$ with $n^a(Y_b)={\delta^a}_b$ as
\begin{eqnarray*}
  Y_{1}&=&\frac{\partial}{\partial u}+H\frac{\partial}{\partial v} \\
  Y_{2}&=&\frac{\partial}{\partial v} \\
  Y_{3}&=&\frac{\partial}{\partial z} \\
  Y_{4}&=&\frac{\partial}{\partial z^{*}}
\end{eqnarray*}
so the metric becomes
\begin{eqnarray}
g=2\big(n^{1}\otimes n^{2}+ n^{2} \otimes n^{1}+n^{3}\otimes n^{4}+n^{4} \otimes n^{3}\big)
\end{eqnarray}
and the dual metric reads
 \begin{eqnarray}
g^{*}=\frac{1}{2}\big(Y_{1}\otimes Y_{2}+ Y_{2} \otimes Y_{1}+Y_{3}\otimes Y_{4}+Y_{4} \otimes Y_{3}\big). \nonumber
\end{eqnarray}
Another convenient co-basis $\{e^{a}\}$ and basis $\{X_{b}\}$ are the orthonormal ones given respectively by
\begin{eqnarray*}
e^{0}&=&(1+H)du-dv \\
e^{1}&=&dz+dz^{*} \\
e^{2}&=&\mathrm{i}(dz^{*}-dz) \\
e^{3}&=&(1-H)du-dv
\end{eqnarray*}
and
\begin{eqnarray*}
X_{0}&=&\frac{1}{2}\big[\frac{\partial}{\partial u}-(1-H)\frac{\partial}{\partial v}\big] \\
X_{1}&=&\frac{1}{2}\big(\frac{\partial}{\partial z}+\frac{\partial}{\partial z^{*} }\big) \\
X_{2}&=&\frac{1}{2\mathrm{i}}\big(\frac{\partial}{\partial z^{*}}-\frac{\partial}{\partial z }\big) \\
X_{3}&=&\frac{1}{2}\big[\frac{\partial}{\partial u}+(1+H)\frac{\partial}{\partial v}\big].
\end{eqnarray*}
So this time the metric is given by
\begin{eqnarray}
g=-e^{0}\otimes e^{0}+\sum_{i=1}^{3}e^{i}\otimes e^{i}
\end{eqnarray}
and the dual metric by
\begin{eqnarray}
g^{*}=-X_{0}\otimes X_{0}+\sum_{i=1}^{3}X_{i}\otimes X_{i}. \nonumber
\end{eqnarray}
The relation between the null co-basis and orthonormal co-basis is
\begin{eqnarray*}
  n^{1}&=&\frac{1}{2}\big(e^{3}+e^{0}\big) \\
  n^{2}&=&\frac{1}{2}\big(e^{3}-e^{0}\big) \\
  n^{3}&=&\frac{1}{2}\big(e^{1}+\mathrm{i}e^{2}\big) \\
  n^{4}&=&\frac{1}{2}\big(e^{1}-\mathrm{i}e^{2}\big)
\end{eqnarray*}
which will be useful later on. The connection 1-forms can be calculated from the torsion-free first structure equation $de^{a}=-{\omega^a}_b \wedge e^b$ and the table follows.
\begin{table}[h]
\centering
\begin{tabular}{c c c c c}
  \hline
  $   {\omega^a}_b  $      &      &     &    & \\ \hline
  $a \setminus b$      & $0$ & $1$ & $2$ &  $3$ \\

  $0$      & $0$ & $\mathcal{A} n^1$ & $\mathcal{B} n^1$ & $0$ \\
  $1$      & $\mathcal{A} n^1$ & $0$ & $0$ & $\mathcal{A} n^1$ \\
  $2$      & $\mathcal{B} n^1$ & $0$ & $0$ & $\mathcal{B} n^1$ \\
  $3$      & $0$ & $-\mathcal{A} n^1$ & $-\mathcal{B} n^1$ & $0$ \\
  \hline
\end{tabular}
\caption{Connection 1-forms}\label{1}
\end{table}

The definitive functions are $\mathcal{A}=\frac{1}{2}(H_z+H_{z^*})$ and $\mathcal{B}=\frac{\mathrm{i}}{2}(H_z-H_{z^*})$. All the covariant derivatives $\nabla_{X_b}e{^a}=-{\omega^a}_c (X_b)e^c$ are easily seen to be
\begin{table}[h]
\centering
\begin{tabular}{c| c c c c}
  \hline
  $   \nabla_{X_b}e{^a}  $      &      &     &    & \\
  $   $               & $e^{0}$ & $e^{1}$ & $e^{2}$ &  $e^{3}$ \\ \hline

  $\nabla_{X_0}$      & $-\frac{1}{2}(\mathcal{A}e^{1}+\mathcal{B}e^{2})$ & $-\mathcal{A} n^1$ & $-\mathcal{B} n^1$ & $\frac{1}{2}(\mathcal{A}e^{1}+\mathcal{B}e^{2})$ \\
  $\nabla_{X_1}$      & $0$ & $0$ & $0$ & $0$ \\
  $\nabla_{X_2}$      & $0$ & $0$ & $0$ & $0$ \\
  $\nabla_{X_3}$      & $-\frac{1}{2}(\mathcal{A}e^{1}+\mathcal{B}e^{2})$ & $-\mathcal{A} n^1$ & $-\mathcal{B} n^1$ & $\frac{1}{2}(\mathcal{A}e^{1}+\mathcal{B}e^{2})$ \\
  \hline
\end{tabular}
\caption{Covariant derivatives of co-basis}\label{2}
\end{table}

It is clear from Table II that $n^1$ is a parallel form i.e. $\nabla_X n^1=0$ for all $X \in \Gamma TM$. The curvature 2-forms can be calculated from the second structure equations ${R^a}_b=d{\omega^a}_b+{\omega^a}_c \wedge {\omega^c}_b$ and it is found that
\begin{eqnarray*}
{R}_{01}&=&{R}_{31}=\frac{1}{4}H_{z{z^*}}(e^0\wedge e^1- e^1\wedge e^3) \\
{R}_{01}&=&{R}_{31}=\frac{1}{4}H_{z{z^*}}(e^0\wedge e^2- e^2\wedge e^3)
\end{eqnarray*}
and from $i_{X_a}{R^a}_b=P_b$ the Ricci 1-forms are $P_0=H_{z{z^*}} n^1=P_3$ and $P_1=0=P_2$ then the scalar curvature is $\mathcal{R}=2H_{z{z^*}}$.
\section{Spinor basis and the inner product}
In this section we use a primitive idempotent $P$ that is a section of the plane wave spacetime's Clifford bundle $C$ to project a minimal left ideal, which is equivalent to a section of the real spinor bundle. After constructing the spinor basis we seek a skew symmetric (symplectic) inner product for spinors that admits the $\xi$ involution as its adjoint. The involutions are the main involutary anti-automorphism $\xi$ or its composition with the automorphism $\eta$ of the Clifford algebra. The definitions of the above maps can easily be given by considering a homogeneous decomposable element of length p of the fibre Clifford algebra that is a decomposable p-form $\alpha=y^1 \wedge y^2 \wedge ... \wedge y^p$ of the exterior fibre algebra at any given point. Then $\alpha^{\xi}=y^p \wedge ... y^2 \wedge y^1=(-1)^{\lfloor p/2\rfloor}\alpha$ and $\alpha^{\eta}=(-1)^{p}\alpha$. It is known that \cite{Talebaoui} the element
 \begin{eqnarray}
P=\frac{1}{2}(1+e^1)\frac{1}{2}(1+e^{30})
\end{eqnarray}
is a primitive idempotent and can naturally be used to build up the desired spinor space. Then a spinor basis $\{\mathbf{b}_i\}$ can be given as
\begin{eqnarray*}
  \mathbf{b}_{1}&=&\frac{1}{4}\big(1+e^{1}-e^{03}+e^{013}\big) \\
  \mathbf{b}_{2}&=&\frac{1}{4}\big(e^{0}+e^{3}+e^{01}-e^{13}\big) \\
  \mathbf{b}_{3}&=&\frac{1}{4}\big(e^{23}-e^{02}+e^{012}+e^{123}\big) \\
  \mathbf{b}_{4}&=&\frac{1}{4}\big(-e^{2}+e^{12}-e^{023}-e^{0123}\big)
\end{eqnarray*}
where $e^{ab...}$ is the abbreviation for $e^{a}\wedge e^{b}\wedge ... $ and one can check that $\mathbf{b}_{i}P=\mathbf{b}_{i}$ for all $i$. We seek an $\mathbb{R}$-skew product $<,>$ with involution $\mathcal{J}\simeq \xi$
\begin{eqnarray*}
  <,>:CP \times CP&\longrightarrow& PCP \\
      \psi,\phi &\mapsto& <\psi,\phi>
\end{eqnarray*}
such that $$<\psi,\phi>=J^{-1}\psi^{\xi}\phi$$ with $J^{\xi}=-J$ and the spinor adjoint is $\overline{\psi}=J^{-1}\psi^{\xi}$. Since $PCP$ is isomorphic to $\mathbb{R}$ as a real algebra and there is no $\mathcal{J}$ that preserves $P$ then there is no involution induced on $\mathbb{R}$. The below relations between the spinor basis will be useful in inner product calculations.
\begin{eqnarray*}
  \mathbf{b}_{1}&=&P \\
  \mathbf{b}_{2}&=&e^{0}P=e^{3}P \\
  \mathbf{b}_{3}&=&e^{23}P=-e^{02}P \\
  \mathbf{b}_{4}&=&-e^{2}P
\end{eqnarray*}
If we choose $J=e^{23}$ then the 6 independent inner products are seen to be
\begin{table}[!ht]
\centering
\begin{tabular}{c| c c c c c}
  \hline
  $<,>   $               & $\mathbf{b}_{1}$ & $\mathbf{b}_{2}$ &  $\mathbf{b}_{3}$ &  $\mathbf{b}_{4}$\\ \hline

  $\mathbf{b}_{1}$      & $0$ & $0$ & $P$ & $0$ \\
  $\mathbf{b}_{2}$      & $0$ & $0$ & $0$ & $P$ \\
  $\mathbf{b}_{3}$      & $-P$ & $0$ & $0$ & $0$ \\
  $\mathbf{b}_{4}$      & $0$ & $-P$ & $0$ & $0$ \\
  \hline
\end{tabular}
\caption{Inner products of spinor basis}\label{3}
\end{table}

It is worth noting that also the necessary property $P^{\xi}=J^{-1}PJ$ is satisfied.

\section{Spinorial covariant derivatives and parallel spinors}
We consider the pseudo-Riemannian connection $\nabla$ on $M$ and we also use the same symbol for differentiating spinor fields, the form given here will also be applicable to the case of non-zero torsion. A covariant derivative $\nabla_X$ on local spinor fields can be defined \cite{Benn and Tucker} by its action on the basis as
\begin{eqnarray}
\nabla_X\mathbf{b}_{i}=\sigma_X \mathbf{b}_{i}
\end{eqnarray}
where $$\sigma_X=\frac{1}{4}\omega_{bc}(X)e^{bc},$$
and satisfying the below properties:
\begin{eqnarray*}
  &(1)&\nabla_{fX}=f\nabla_X     ;   \forall f\in\mathcal{F}(M),  \\
  &(2)&\nabla_X(a\psi)=(\nabla_Xa)\psi+a\nabla_X\psi    ;\forall a\in\Gamma C(M), \forall \psi\in\Gamma S(M),  \\
  &(3)& <\nabla_X\psi,\phi>+<\psi,\nabla_X\phi>=X<\psi,\phi>.
\end{eqnarray*}
By the help of Table I one finds $$\sigma_{X_0}=\sigma_{X_3}=\frac{1}{2}(\mathcal{A}e^1+\mathcal{B}e^2)\wedge n^1=\frac{1}{2}(\mathcal{A}e^1+\mathcal{B}e^2)n^1$$ and the left Clifford action of $n^1$ on the base spinors is
\begin{eqnarray*}
  n^1\mathbf{b}_{1}&=&\mathbf{b}_{2} \\
  n^1\mathbf{b}_{2}&=&0 \\
  n^1\mathbf{b}_{3}&=&0 \\
  n^1\mathbf{b}_{4}&=&\mathbf{b}_{3}
\end{eqnarray*}
as a result we see that $n^1$ belongs to the intersection of the distribution of null spaces of $\mathbf{b}_{2}$ and $\mathbf{b}_{3}$ i.e. $n^1\in \mathcal{N}_{\mathbf{b}_{2}}\cap \mathcal{N}_{\mathbf{b}_{3}}$; we also note that $n^2\in \mathcal{N}_{\mathbf{b}_{1}}\cap \mathcal{N}_{\mathbf{b}_{4}}$. Then the covariant derivatives of the spinor basis are given below.
\begin{table}[h]
\centering
\begin{tabular}{c| c c c c}
  \hline
  $   \nabla_{X_a}\mathbf{b}_{i}  $      &      &     &    & \\
  $   $               & $\mathbf{b}_{1}$ & $\mathbf{b}_{2}$ & $\mathbf{b}_{3}$ &  $\mathbf{b}_{4}$ \\ \hline

  $\nabla_{X_0}$      & $-\frac{1}{2}(\mathcal{A}\mathbf{b}_{2}-\mathcal{B}\mathbf{b}_{3})$ & $0$ & $0$ & $\frac{1}{2}(\mathcal{B}\mathbf{b}_{2}+\mathcal{A}\mathbf{b}_{3})$ \\
  $\nabla_{X_1}$      & $0$ & $0$ & $0$ & $0$ \\
  $\nabla_{X_2}$      & $0$ & $0$ & $0$ & $0$ \\
  $\nabla_{X_3}$      & $-\frac{1}{2}(\mathcal{A}\mathbf{b}_{2}-\mathcal{B}\mathbf{b}_{3})$ & $0$ & $0$ & $\frac{1}{2}(\mathcal{B}\mathbf{b}_{2}+\mathcal{A}\mathbf{b}_{3})$ \\
  \hline
\end{tabular}
\caption{Covariant derivatives of spinor basis}\label{4}
\end{table}

Table IV shows that $\mathbf{b}_{2}$ and $\mathbf{b}_{3}$ are parallel spinors.

\section{Massless Dirac spinors and the Rarita-Schwinger solutions}
I.The two parallel spinors can be used to build up massless \cite{Benn and Tucker} and massive Dirac spinors \cite{Talebaoui} in plane wave spacetimes. If $h(u)$ and $g(u)$ are arbitrary functions of the coordinate $u$ then $$\psi=h(u)\mathbf{b}_{2}+g(u)\mathbf{b}_{3}$$ is a massless Dirac spinor satisfying $\displaystyle{\not}D\psi=0$. The Dirac operator is defined by the Clifford contraction $\displaystyle{\not}D:=e^a\nabla_{X_a}$. Furthermore it is easily seen that there are two more massless Dirac spinors. $e^0 \nabla_{X_0}\mathbf{b}_{1}=\frac{1}{2}(\mathcal{A}\mathbf{b}_{1}-\mathcal{B}\mathbf{b}_{4})$ and $e^3 \nabla_{X_3}\mathbf{b}_{1}=-\frac{1}{2}(\mathcal{A}\mathbf{b}_{1}-\mathcal{B}\mathbf{b}_{4})$ and since $e^i\nabla_{X_i}\mathbf{b}_{1}=0$ for $i=1,2$ then $$\displaystyle{\not}D\mathbf{b}_{1}=0.$$ Similarly $e^0 \nabla_{X_0}\mathbf{b}_{4}=-\frac{1}{2}(\mathcal{B}\mathbf{b}_{1}+\mathcal{A}\mathbf{b}_{4})$ and $e^3 \nabla_{X_3}\mathbf{b}_{4}=\frac{1}{2}(\mathcal{B}\mathbf{b}_{1}+\mathcal{A}\mathbf{b}_{4})$ and since $e^i\nabla_{X_i}\mathbf{b}_{4}=0$ for $i=1,2$ then also $$\displaystyle{\not}D\mathbf{b}_{4}=0.$$\\

II. A spinor-valued 1-form $\Psi=\psi_a\otimes e^a$ is a massive Rarita-Schwinger field if $e^a\psi_a=0$ and $\displaystyle{\not}D\Psi=m\Psi$ such that the spin-3/2 covariant derivative is defined as $$\nabla_X\Psi=\nabla_X \psi_a \otimes e^a+\psi_a \otimes \nabla_Xe^a.$$ We propose that $\Psi_k:=\nabla_{X_a}\mathbf{b}_{k}\otimes e^a$ for $k=1,4$ satisfies the Rarita-Schwinger equations the mass of which is determined by the analytical properties of the profile function $H$. We here will start with the proof for $k=1$ and then proceed for $k=4$ case. For $k=1$ our spin tensor clearly is $\Psi_1=-\frac{1}{2}(\mathcal{A} \mathbf{b}_{2}-\mathcal{B}\mathbf{b}_{3})\otimes (e^0+e^3)$ then $\nabla_{X_a}\Psi_1=\big(X_a(\mathcal{B}) \mathbf{b}_{3}-X_a(\mathcal{A})\mathbf{b}_{2}\big)\otimes n^1$, so $\displaystyle{\not}D\Psi_1=(d\mathcal{B}.\mathbf{b}_{3}-d\mathcal{A}.\mathbf{b}_{2})\otimes n^1$. Remembering that $\mathcal{A}=\frac{1}{2}(H_z+H_{z^*})$ and $\mathcal{B}=\frac{\mathrm{i}}{2}(H_z-H_{z^*})$ then $d\mathcal{A}=\frac{1}{2}\big[(H_{zz}+H_{z^*z})dz+(H_{zz^*}+H_{z^*z^*})dz^*+(H_{zu}+H_{z^*u})du\big]$ and $d\mathcal{B}=\frac{\mathrm{i}}{2}\big[(H_{zz}-H_{z^*z})dz+(H_{zz^*}-H_{z^*z^*})dz^*+(H_{zu}-H_{z^*u})du\big].$ The left actions of these on $\mathbf{b}_{2}$ and $\mathbf{b}_{3}$ respectively are
\begin{eqnarray}
  d\mathcal{A}.\mathbf{b}_{2}&=&-\frac{1}{4}\big[(H_{zz}+H_{z^*z})(\mathbf{b}_{2}-\mathrm{i}\mathbf{b}_{3})+(H_{zz^*}+H_{z^*z^*})(\mathbf{b}_{2}+\mathrm{i}\mathbf{b}_{3})\big] \\ \nonumber
  d\mathcal{B}.\mathbf{b}_{3}&=&\frac{1}{4}\big[(H_{zz}-H_{z^*z})(\mathbf{b}_{2}-\mathrm{i}\mathbf{b}_{3})-(H_{zz^*}-H_{z^*z^*})(\mathbf{b}_{2}+\mathrm{i}\mathbf{b}_{3})\big]
\end{eqnarray}
and finally
\begin{eqnarray}
\displaystyle{\not}D\Psi_1=\frac{1}{2}\big[(H_{zz}+H_{z^*z^*})\mathbf{b}_{2}-\mathrm{i}(H_{zz}-H_{z^*z^*})\mathbf{b}_{3}\big]\otimes n^1.
\end{eqnarray}
So $$\Psi_1=-\frac{1}{2}\big[(H_z+H_{z^*}) \mathbf{b}_{2}-\mathrm{i}(H_z-H_{z^*})\mathbf{b}_{3}\big]\otimes n^1$$ is a Rarita-Schwinger field with mass $m$ if and only if
\begin{eqnarray}
(H_{zz}\pm H_{z^*z^*})=m(H_{z}\pm H_{z^*}).
\end{eqnarray}
An analytical solution of this kind is found to be $H(u,z,z^*)=f(u)e^{m(z+z^*)}$ which certainly annihilates the $\mathbf{b}_{3}$ part of the spin tensor.\\

For $k=4$ our spin tensor is $\Psi_4=\frac{1}{2}(\mathcal{B} \mathbf{b}_{2}+\mathcal{A}\mathbf{b}_{3})\otimes (e^0+e^3)$ then $\nabla_{X_a}\Psi_4=\big(X_a(\mathcal{B}) \mathbf{b}_{2}+X_a(\mathcal{A})\mathbf{b}_{3}\big)\otimes n^1$, so $\displaystyle{\not}D\Psi_4=(d\mathcal{B}.\mathbf{b}_{2}+d\mathcal{A}.\mathbf{b}_{3})\otimes n^1$. $d\mathcal{A}=\frac{1}{2}\big[(H_{zz}+H_{z^*z})dz+(H_{zz^*}+H_{z^*z^*})dz^*+(H_{zu}+H_{z^*u})du\big]$ and $d\mathcal{B}=\frac{\mathrm{i}}{2}\big[(H_{zz}-H_{z^*z})dz+(H_{zz^*}-H_{z^*z^*})dz^*+(H_{zu}-H_{z^*u})du\big].$ The left actions of these on $\mathbf{b}_{3}$ and $\mathbf{b}_{2}$ respectively are
\begin{eqnarray}
  d\mathcal{A}.\mathbf{b}_{3}&=&\frac{1}{4}\big[(H_{zz}+H_{z^*z})(\mathbf{b}_{3}+\mathrm{i}\mathbf{b}_{2})+(H_{zz^*}+H_{z^*z^*})(\mathbf{b}_{3}-\mathrm{i}\mathbf{b}_{2})\big] \\ \nonumber
  d\mathcal{B}.\mathbf{b}_{2}&=&\frac{1}{4}\big[-(H_{zz}-H_{z^*z})(\mathbf{b}_{3}+\mathrm{i}\mathbf{b}_{2})+(H_{zz^*}-H_{z^*z^*})(\mathbf{b}_{3}-\mathrm{i}\mathbf{b}_{2})\big]
\end{eqnarray}
and finally
\begin{eqnarray}
\displaystyle{\not}D\Psi_4=H_{z^*z^*}\mathbf{b}_{3}\otimes n^1.
\end{eqnarray}
So $$\Psi_4=\frac{1}{2}\big[\mathrm{i}(H_z-H_{z^*})\mathbf{b}_{2}+(H_z+H_{z^*}) \mathbf{b}_{3}\big]\otimes n^1$$ is a Rarita-Schwinger field with mass $m$ if and only if
\begin{eqnarray}
H_{z}=H_{z^*} , H_{zz}=mH_{z}.
\end{eqnarray}
However, since the spin-3/2 norms of these solutions are vanishing, this means that the physical solutions should be massless.
\section{Symmetry considerations}
We here assume a first order symmetry operator \cite{Benn Kress2}
\begin{eqnarray}
L_{\omega}=e^a \omega \nabla_{X_a}+\frac{p}{p+1}d\omega-\frac{n-p}{n-p+1}d^{\dag}\omega
\end{eqnarray}
that maps $\mathbf{b}_{1}$ to $\mathbf{b}_{4}$ and which has degree 1. Since $L_{*\omega}$ is also a symmetry operator the degree 3 case will be a result of the degree 1 case, but the degree 2 case will be analyzed elsewhere. Here $d^{\dag}$ is the co-derivative and for the pseudo-Riemannian spacetimes $d=e^a \wedge \nabla_{X_a}$ and $d^{\dag}=-i_{X^a}\nabla_{X_a}$. Let us calculate $d\omega$, $d^{\dag}\omega$ and $Q_{\omega}:=e^a \omega \nabla_{X_a}$ for a general 1-form $\omega$. If we write $$\omega=\alpha e^0+\beta e^1+ \gamma e^2+\delta e^3$$ then
\begin{eqnarray}
d\omega&=&\frac{\alpha_{u}}{2}e^{30}-\beta_{u}e^1n^1-\gamma_{u}e^2n^1-\frac{\delta_{u}}{2}e^{30}-\alpha_{z}e^0n^3-\frac{\mathrm{i}\beta_{z}}{2}e^{12}-\frac{\gamma_{z}}{2}e^{21}\\ \nonumber
&-&\delta_{z}e^3n^3-\alpha_{z^*}e^0n^4+\frac{\mathrm{i}\beta_{z^*}}{2}e^{12}-\frac{\gamma_{z^*}}{2}e^{21}-\delta_{z^*}e^3n^4 \nonumber
\end{eqnarray}
which can more compactly be written as
\begin{eqnarray}
d\omega&=&\frac{(\alpha_{u}-\delta_{u})}{2}e^{30}-\frac{\mathrm{i}(\beta_z-\beta_{z^*})-(\gamma_z+\gamma_{z^*})}{2}e^{12}\\ \nonumber&-&(\beta_u e^1+\gamma_u e^2)n^1-(\alpha_z e^0+ \delta_z e^3)n^3-(\alpha_{z^*}e^0+\delta_{z^*}e^3)n^4.
\end{eqnarray}
The co-derivative part heavily simplifies because
\begin{eqnarray*}
  i_{X^a}\nabla_{X_a}e^0&=&0 \\
  i_{X^a}\nabla_{X_a}e^1&=&-\mathcal{A}(i_{X^0}n^1+i_{X^3}n^1)=0 \\
  i_{X^a}\nabla_{X_a}e^2&=&-\mathcal{B}(i_{X^0}n^1+i_{X^3}n^1)=0 \\
  i_{X^a}\nabla_{X_a}e^3&=&0
\end{eqnarray*}
so
\begin{eqnarray}
d^{\dag}\omega=\frac{1}{2}(\alpha_u-(\beta_z+\beta_{z^*})+\mathrm{i}(\gamma_{z^*}-\gamma_z)-\delta_u).
\end{eqnarray}
The $Q_{\omega}\mathbf{b}_{1}$ can be calculated as
\begin{eqnarray}
Q_{\omega}\mathbf{b}_{1}=\mathcal{A}(\alpha-\delta)\mathbf{b}_{2}-\mathcal{B}(\alpha-\delta)\mathbf{b}_{3}.
\end{eqnarray}
Finally
\begin{eqnarray}
L_{\omega}\mathbf{b}_{1}&=&\mathcal{A}(\alpha-\delta)\mathbf{b}_{2}-\mathcal{B}(\alpha-\delta)\mathbf{b}_{3}+\frac{1}{4}(\alpha_u-\delta_u)\mathbf{b}_{1}+\frac{1}{4}(2\beta_u-\alpha_z-\delta_z-\alpha_{z^*}-\delta_{z^*})\mathbf{b}_{2}\\ \nonumber
&+&\frac{1}{4}(-2\gamma_u+\mathrm{i}(\alpha_z+\delta_z-\alpha_{z^*}-\delta_{z^*}))\mathbf{b}_{3}+\frac{1}{4}(\gamma_z+\gamma_{z^*}-\mathrm{i}(\beta_z-\beta_{z^*}))\mathbf{b}_{4} \\ \nonumber
&-&\frac{3}{8}(\alpha_u-\beta_z-\beta_{z*}+\mathrm{i}(\gamma_{z^*}-\gamma_z)-\delta_u)\mathbf{b}_{2}=\mathbf{b}_{4}
\end{eqnarray}
which gives
\begin{eqnarray}
  \big[\delta_u-\alpha_u+3\beta_z+3\beta_{z^*}-3\mathrm{i}(\gamma_{z^*}-\gamma_z)\big]&=&0 \\ \nonumber
  \big[4\mathcal{A}(\alpha-\delta)+2\beta_u-\alpha_z-\delta_z-\alpha_{z^*}-\delta_{z^*}\big]&=&0 \\ \nonumber
  \big[-4\mathcal{B}(\alpha-\delta)+2\gamma_u+\mathrm{i}(\alpha_z+\delta_z-\alpha_{z^*}-\delta_{z^*})\big]&=&0 \\ \nonumber
  \big[\gamma_z+\gamma_{z^*}-\mathrm{i}(\beta_z-\beta_{z^*})\big]&=&1.
\end{eqnarray}
This set of first order linear partial differential equations determine the general solution, but if we assume that $\alpha=\delta$ then a special solution for the symmetry covector is found to be
\begin{eqnarray}
\omega=2(z+z^*)[du-dv+i(dz^{*}-dz)]+2u(dz^{*}+dz).
\end{eqnarray}\\

One can show that the only non-vanishing singled bilinear corresponding to the standard spinor basis is $$\mathbf{b}_{2}\overline{\mathbf{b}_{2}}=-2(1-e^1)n^{1}e^{2},$$ generated by one of the parallel spinors, namely $\mathbf{b}_{2}$; so by defining $\Omega_p:= (\mathbf{b}_{2}\overline{\mathbf{b}_{2}})_p;\, p=2,3$ then it is found that $\Omega_3=2 e^1 n^1 e^2$ and $\Omega_2=2 e^2 n^1$. Together with $\Omega_1:=2n^1$, these constitute a set of null parallel forms over the plane wave spacetime. Since the null direction is determined by the vector field $\widetilde{n^1}$, these forms can be interpreted as conserved quantities of parallel propagating null (p-1)-branes along $\widetilde{n^1}$. This set of null parallel forms are related to each other by the following algebraic relations $$i_{X_1}\Omega_3=\Omega_2\,,\,i_{X_2}\Omega_2=\Omega_1.$$

\section{Conclusion}
We first considered solutions of the massless Dirac equation in plane wave spacetimes and found two solutions apart from the ones generated by parallel spinors. Later on we focused on the solutions of Rarita-Schwinger equation where we have built up spin tensors out of the massless Dirac spinors that we obtained before and reached massive Rarita-Schwinger solutions. The mass parameter of the field come out as a consequence of some constraint equations containing the profile function of the plane wave spacetime; which also included the massless limit. This limit is presumed to be more physical than the massive case because of the self inner products of the Rarita-Schwinger fields that give null spinor-valued one forms. Our next consideration was the first order symmetry operators of the massless Dirac equation transforming the solutions that we have obtained and at least a symmetry generating covector field is calculated. It is interesting to note that since our solutions $\Psi=\psi_a \otimes e^a$ satisfy $P^a.\psi_a=0$, the plane wave spacetime class is convenient for the quantization of spin-3/2 fields non-minimally coupled to the gravitational field \cite{Hack and Makedonski}.\\

 Another supporting result for our solutions in classical gravity is that the last condition also appears as an algebraic constraint equation in the search for creation of spin-$3/2$ particles in strong gravitational fields, while the minimal coupling to electromagnetic interactions are turned off \cite{Gibbons}. It is also believed that an intuitive understanding of spinor-valued gauge fields requires the construction of solutions to the definitive equations in classical backgrounds. For this aim our solutions can be thought as the solutions for the limiting equations of coupled Einstein-Rarita-Schwinger theory which decouple for a convenient gauge fixing of the fermionic sector \cite{Schima and Kasuya}; though this has to be worked out separately. In the last reference \cite{Schima and Kasuya} the non-triviality of the Rarita-Schwinger field solutions are thought to be the sources for the non-triviality of the topology for the classical background; which for our case indicates possible different topologies for the plane wave geometry. Although the Penrose's result for the correspondence between the conserved charges of spin-$3/2$ fields and flat twistor space \cite{Penrose 3/2} breaks down for curved spacetimes \cite{Silva-Ortigoza}, and since Killing spinors are particular twistors that can be use to generate Rarita-Schwinger solutions in constant scalar curvature spacetimes \cite{Acik 2018}, another relation between twistors and spin-$3/2$ conserved charges can then be found. Gauged versions of the Rarita-Schwinger solutions can also be considered as future work, though it is known that the minimally gauged massless Rarita-Schwinger field yields a consistent classical theory \cite{Adler}.

\appendix
\section{Tensor spinors}
Independent of the dimension and signature, the higher dimensional half-integer representations of the spin group can be constructed by tensoring the tensors over an inner product space with the spinors of the associated Clifford algebra. These \textit{tensor spinors} could also be thought of as \textit{spinor-valued tensors} \cite{Benn and Tucker}. A first example for tensor spinors can be given by \textit{spinor-valued $1$-forms} over a pseudo-Riemannian spacetime assumed as a spin manifold $(M,g)$. If ${e^a}$ is any co-frame and ${\mathbf{b}_i}$ is any standard spinor frame then one can write $$\Psi=\psi_a \otimes e^a$$ or equivalently $$\Psi=\mathbf{b}_i \otimes \psi^i$$ where the spinors $\psi_a$'s and $1$-forms $\psi^i$'s are defined respectively as $\psi_a=\mathbf{b}_i \psi_a^i$ and $\psi^i=\psi_a^i e^a$.\\

 In four spacetime dimensions the complex Clifford algebra is isomorphic to the algebra of $4\times4$ complex matrices $\mathbb{C}(4)$ and the even algebra is isomorphic to $\mathbb{C}(2)\oplus \mathbb{C}(2)$ whereas the real subalgebra is isomorphic to $\mathbb{R}(4)$. The irreducible representations of these algebras are carried respectively by complex (Dirac) spinors, complex semi-spinors (Weyl/chiral spinors) and real (Majorana) spinors. Since $z^2=-1$ for Lorentzian signature we can choose $\psi_a$'s as chiral spinors satisfying $iz\psi_a=\psi_a$, hence they carry irreducible representations of the spin group $SL(2,\mathbb{C})$. $1$-forms Clifford anti-commute with the volume form $z$, then they can be thought of as $(1,1)$ tensors on spinors changing the semi-spinor components under their left Clifford action. Consequently we may regard spinor-valued $1$-forms as bidegree (1,2) tensors on spinor fields. If $u$ and $v$ are two chiral spinors lying in the same component with $\psi_a$, then one can define $$\Psi(u,v):=(u,\psi_a)e^a.v .$$ It turns out that irreducible $SL(2,\mathbb{C})$ representations are carried by spinor-valued $1$-forms that are totally symmetric in their covariant and contravariant arguments separately \cite{Penrose Rindler II}. This is equivalent to saying that the spinor-valued $1$-form is an irreducible spin tensor if it is ''traceless'': $$e^a.\psi_a=0.$$ For example if we assume that $\psi$ satisfies the Weyl equation $\displaystyle{\not}D\psi=0$ then $\Psi$ is traceless if $\psi_a:=\nabla_{X_a}\psi.$\\

 The covariant derivative $\nabla_X$ on spinors and tensors can be extended by the Leibniz rule to a covariant derivative on spinor-valued $1$-forms, also denoted by $\nabla_X$, as $$\nabla_X\Psi=\nabla_X \psi_a \otimes e^a+\psi_a \otimes \nabla_Xe^a.$$ A higher half-integral spin representation of Clifford algebra on spinor-valued $1$-forms can be defined by $$a.\Psi:=(a.\psi_b)\otimes e^b.$$ Then with the tracelessness condition the Dirac-like equation $$\displaystyle{\not}D\Psi=m \Psi$$ constitute the spin $3/2$ Rarita-Schwinger equation. A spinor-valued $p$-form can likewise be defined as $$\Psi:=\psi_{I(p)}\otimes e^{I(p)}$$ where $I(p)$ is an ordered multi-index for a $p$-form basis. If $a$ is any $q$-form and $\Psi$ a spinor-valued $p$-form then a spinor-valued $p+q$-form can be build up from them as $$a\wedge \Psi=\psi_{I(p)}\otimes a\wedge e^{I(p)}.$$ The covariant derivative is again extended to this sector by the Leibniz rule and the spinor covariant exterior derivative D maps spinor-valued $p$-forms to spinor-valued $p+1$-forms: $$D\Psi=e^a \wedge \nabla_{X_a} \Psi.$$ If $\{\mathbf{b}_i\}$ is a standard spinor frame and $\{\psi^i=\psi^i_{I(p)} e^{I(p)}\}$ a set of $p$-forms then $\Psi$ may be expanded as $$\Psi=\mathbf{b}_i \otimes \psi^i$$ and equivalently the spinor exterior covariant derivative can be written as $$D\Psi=d\Psi+\omega.\Psi$$ where $d\Psi:=\mathbf{b}_i \otimes d\psi^i$ and $\omega$ is the Clifford-valued connection $1$-form $\omega:=\frac{1}{4}e^{pq} \otimes \omega_{pq}$. If $n^{A(q)}=\sum_{r}(n^{A(q)})_{J(r)} e^{J(r)}$'s are arbitrary Clifford forms, a Clifford-valued $q$-form $N=n^{A(q)} \otimes e_{A(q)}$ left acts on $\Psi$ as $$N.\Psi=n^{A(q)}.\psi_{I(p)} \otimes e_{A(q)}\wedge e^{I(p)}.$$ Another possible description for the Rarita-Schwinger equation can be given as $$e*D\Psi=0$$ where $e:=e^a \otimes e_a$.
The spin-3/2 inner product of two spinor-valued one forms for $\Psi=\psi_a \otimes e^a$ and $\Phi=\phi_b \otimes e^b$ is $$<\Psi,\Phi>=<\psi_a,\phi_b>g(e^a,e^b).$$

 \end{document}